\documentclass[openacc]{rsproca_new}%%%%where rsproca is the template name

\begin{document}

%%%% Article title to be placed here
\title{Selective and tunable excitation of topological non-Hermitian skin modes }%and non-Hermitian self-healing}

\author{%%%% Author details
Stefano Longhi$^{1,2}$}

%%%%%%%%% Insert author address here
\address{$^{1}$ Dipartimento di Fisica, Politecnico di Milano, Piazza Leonardo da Vinci 32, I-20133 Milano (Italy)\\
$^{2}$ IFISC (UIB-CSIC), Instituto de Fisica Interdisciplinar y Sistemas Complejos, E-07122 Palma de Mallorca (Spain) }

%%%% Subject entries to be placed here %%%%
\subject{Non-Hermitian Physics, Topological matter}

%%%% Keyword entries to be placed here %%%%

%%%% Insert corresponding author and its email address}
\corres{Stefano Longhi\\
\email{stefano.longhi@polimi.it}}

%%%% Abstract text to be placed here %%%%%%%%%%%%
\begin{abstract}
Non-Hermitian lattices under semi-infinite boundary conditions sustain an extensive number of exponentially-localized states, dubbed non-Hermitian skin modes. Such states can be predicted from the nontrivial topology of the energy spectrum under periodic boundary conditions via a bulk-edge correspondence. However, the selective excitation of the system in one among the infinitely-many topological skin edge states is challenging both from practical and conceptual viewpoints. In fact, in any realistic system with a finite lattice size  most of skin edge states collapse and become metastable states. Here we suggest a route toward the selective and tunable excitation of topological skin edge states which avoids the collapse problem by emulating semi-infinite lattice boundaries via tailored on-site potentials at the edges of a finite lattice. We illustrate such a strategy by considering a non-Hermitian topological interface obtained by connecting two Hatano-Nelson chains with opposite imaginary gauge fields, which is amenable for a full analytical treatment.
\end{abstract}
%%%%%%%%%%%%%%%%%%%%%%%%%%%

%%%%%%%%%% Insert the texts which can accomdate on firstpage in the tag "fmtext" %%%%%

\begin{fmtext}
\section{Introduction}
 Non-Hermitian topological physics \cite{A1,A2,A3,A4,A5} is attracting a considerable interest since the past few years, with a wealth of novel phenomena which do not have any counterpart in corresponding Hermitian topological systems (see \cite{A6,A7,A8,A8b,A9,A10,A11,A12,A13,A14,A15,A16,A17,A18,A19,A20,A21,A22,A23,A24,A25,A26,A27,A28,A29,A30,A31,A32,A33,A34,A35,A36,A37,A38,A39,A40,A41,A42,A43,A44,A45,A46,A47} and references therein). The ability to experimentally implement and control non-Hermiticity using synthetic lattices has been demonstrated using different physical platforms ranging from photonic \cite{A34,A37,A38,A46,A47}, acoustic \cite{A13,A44} and micro
 mechanical \cite{A32,A33} systems  to topolectrical circuits \cite{A35,A36}.
 
\end{fmtext}
\maketitle
 A central result of topological materials is the bulk-edge correspondence:  when two materials with different Bloch bulk topological invariants are interfaced, localized edge
states emerge with energies that lie within the energy gap of the surrounding bulk media \cite{A48,A49}. However, such a rather universal result is challenged
in  non-Hermitian systems, where
 the bulk-boundary correspondence apparently breaks
down \cite{A6,A9,A10} and localization of an extensive number of bulk states 
at the edges is observed, a phenomenon dubbed the non-Hermitian skin effect \cite{A9,A10,A14}. In general a non-Hermitian lattice possesses two different types of topological edges states: 
 the conventional ones that have a Hermitian counterpart, and the non-Hermitian skin modes without any Hermitian counterpart. This means that two different bulk-boundary correspondences can be established in non-Hermitian topological systems \cite{A28}. 
The former one relates edge states to the wave function bulk topological invariants, however the band topology should
be described by the non-Bloch band theory, where the Bloch wave vector is complex and varies over the generalized Brillouin zone \cite{A17,A18,A28}. The latter bulk-boundary correspondence relates the non-Hermitian skin modes to the spectral topology of the Bloch Hamiltonian over the ordinary Brillouin zone \cite{A25,A29}.  A nontrivial spectral topology, as measured by a non-vanishing winding number $W$, results in the existence of topological skin modes \cite{A1,A25,A29,A41}. Experimental demonstrations of skin edge states have been reported in recent works\cite{A32,A34,A35,A36,A37}. In particular, localization of the excitation at the interface separating two non-Hermitian lattices with different spectral topological winding numbers has been observed, even in the absence of ordinary edge states, indicating the collapse of all eigenstates at the interface \cite{A34,A37}. However a rather challenging and still open question is to find a way to selective excite a single skin mode, among the infinitely many ones sustained by the system under semi-infinite boundaries. The difficulty is not only practical, due to the complication of preparing the system in a pure skin edge state, but also conceptual. To clarify this point, let us consider as as example a single-band one-dimensional lattice with a nontrivial spectral topology, i.e. with a Bloch Hamiltonian $H(k)$ whose spectrum describes a closed loop in complex energy plane as $k$ spans the ordinary Brillouin zone. A prototypal example is provided by the clean Hatano-Nelson model \cite{A1,A50}, where $H(k)$ describes an ellipse in complex energy plane. According to the bulk-edge correspondence, for any complex energy $E$ internal to the loop with a negative winding number $W(E)<0$ there exist $|W(E)|$ edge states localized at the left boundary of the lattice \cite{A1,A25}. However, as soon as a right boundary is introduced, the edge states collapse and only those with complex energies belonging to the Hamiltonian $H(k)$, with $k$ varying over the generalized Brillouin zone, survive \cite{A1}. This result makes the physical relevance of such skin edge modes questionable, suggesting that most of them are metastable states ({\it quasi-edge states}) that can be observed only transiently for a short time \cite{A1}.\\  
In this work we show that skin edge modes can be selectively and stably excited in a finite-size non-Hermitian lattice from the free dynamical evolution of the system under initial single-site edge excitation, provided that energy at the edges is judiciously supplied to the system. Basically, complex on-site potentials are added at the edges of a finite-size interface,  so as to emulate semi-infinite boundary conditions to stabilize a target skin edge state. We illustrate this major result by considering skin edge modes in the Hatano-Nelson model \cite{A1,A25,A50,A51,A52}, since it provides the simplest system displaying the non-Hermitian skin effect amenable for a full analytical treatment and accessible to the experiments using synthetic lattices. Specifically, we show selective and tunable excitation of skin edge states at the interface (domain wall) between two Hatano-Nelson chains with opposite imaginary gauge fields \cite{A45}. 

\section{Skin modes in the Hatano-Nelson topological interface}
 The clean Hatano-Nelson model describes the hopping dynamics of a quantum particle on a one dimensional lattice with asymmetry in the left/right hopping amplitudes induced by an imaginary gauge field \cite{A1,A50,A52}. Assuming rather generally an inhomogeneous imaginary gauge field $h=h(n)$, the Schr\"odinger equation for the wave amplitudes $\psi_n$ at the various lattice sites reads \cite{A45}
\begin{equation}
i \frac{d \psi_n}{dt}= \Delta \left\{ \exp [ h(n+1)] \psi_{n+1}+ \exp [-h(n)] \psi_{n-1} \right\} \label{noo}
\end{equation}
where $\Delta \exp(\pm h)$ are the left/right hopping amplitudes.\\ 
Let us first briefly recall the topological properties of the Hatano-Nelson model in the homogeneous case $h(n)=h \; {\rm constant}$ \cite{A1,A25}. Under periodic boundary conditions (PBC), with the Ansatz $\psi_n=\exp [ ikn-i H(k) t]$ the Hamiltonian in Bloch space  reads $H(k)=2 \Delta \cosh(h+ik)$, where $-\pi \leq k < \pi$ is the Bloch wave number. The corresponding energy spectrum describes a closed loop (an ellipse) in complex plane [Fig.1(a)], defined by the equation
\begin{equation}
\left( 
\frac{{\rm Re}(H)}{2 \Delta \cosh h}
\right)^2+\left( 
\frac{{\rm Im}(H)}{ 2 \Delta \sinh h}
\right)^2=1. \label{uff}
\end{equation}
The ellipse is travelled clockwise for $h<0$, and counter-clockwise for $h>0$.
In the bulk of the lattice, a backward (forward) drift current is observed for $h>0$ ($h<0$) with a velocity $v=2 \Delta \sinh |h|$ \cite{A51}. The spectral topology is described by the winding number $W(E)$ \cite{A1,A25}
\begin{equation}
W(E)=\frac{1}{2 \pi i} \int_{-\pi}^{\pi} dk \log \left\{ H(k)-E \right\}
\end{equation}
for a given complex energy $E$. Clearly, one has $W(E)=0$ when $E$ is external to the ellipse, while $W(E)=h/ |h|= \pm 1$ when $E$ is internal to the ellipse. A bulk-edge correspondence can be established for the skin modes under the semi-infinite boundary conditions (SIBC) \cite{A1,A25}: for any complex energy $E$, there are exactly $|W(E)|$ skin edge eigenstates with eigenenergy $E$, which are exponentially localized at the left edge  of the semi-infinite lattice for $W(E)<0$, or at the right edge of the semi-infinite lattice for $W(E)>0$. Therefore, under SIBC the continuous set of the skin edge eigenstates entirely fills the interior of the ellipse. 
Under open boundary conditions (OBC), the energy spectrum collapses to the segment $(-\Delta,\Delta)$ on the real axis [Fig.1(a)], and thus only a subset of skin edge modes on an open line survive \cite{A1,A25}.\\  
Let us now consider the Hatano-Nelson topological interface \cite{A45}, where two Hatano-Nelson lattices with different values of the imaginary gauge field are connected [Fig.1(b)]. Specifically, we assume equal but opposite values of the gauge fields by letting in Eq.(\ref{noo}) $h(n)=-h<0$ for $n \leq 0$ and $h(n)=h>0$ for $n>0$. Clearly, for each of the two Hatano-Nelson chains the PBC energy spectrum is described by the same ellipse in complex energy plane [Eq.(\ref{uff})],  but the two ellipses are travelled in opposite directions [Fig.1(b)], leading to opposite values of the winding numbers $W_1(E)=-W_2(E)=1$ for any energy $E$ in the interior of the ellipse while $W_1(E)=W_2(E)=0$ for any energy $E$ in the exterior of the ellipse. According to the bulk-boundary correspondence, exponentially-localized topological skin modes do exist at the interface, for infinitely-extended chains at both sides, at any energy $E$ in the interior of the ellipse, corresponding to different topological numbers $W_1 \neq W_2$ \cite{A45}. The explicit form of the interface skin modes can be readily obtained and reads 
 \begin{equation}
 \psi_n= \exp( ikn+\mu n -h |n|-iEt) \label{palle}
 \end{equation}
where $-\pi \leq k \leq \pi$, $-h <\mu< h$ for localization, and % $A$ is an arbitrary normalization constant and 
\begin{equation}
E=2 \Delta \cosh (\mu+ik)
\end{equation} 
\begin{figure}[!h]
\centering\includegraphics[width=4.5in]{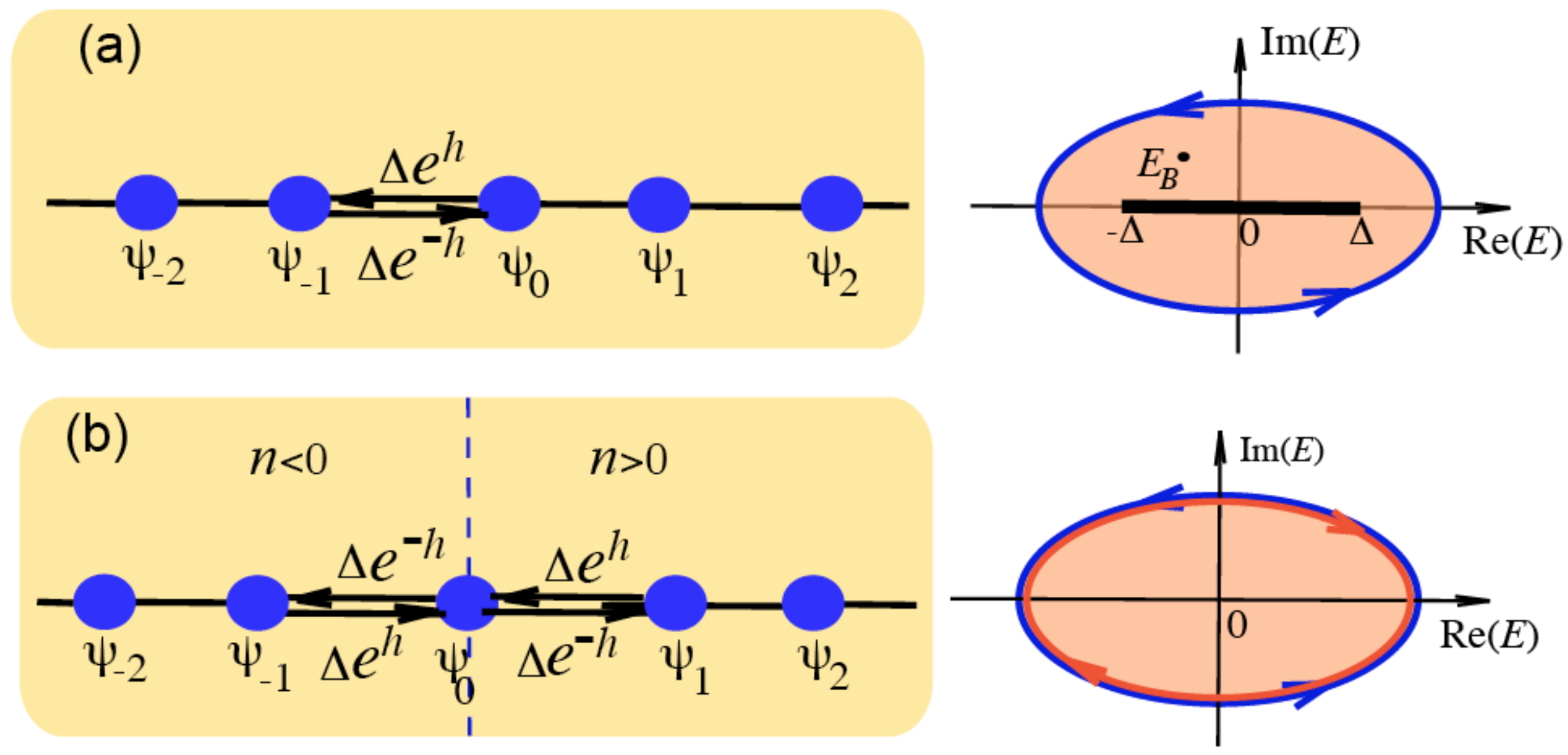}
%%% where xxxxxx name represents "figurename.eps"
\caption{(a) Schematic of the Hatano-Nelson lattice with asymmetric hopping amplitudes $\Delta \exp ( \pm h)$. The PBC energy spectrum is an ellipse [Eq.(\ref{uff})] in complex energy plane (right panel), which is traveled clockwise for $h<0$ and counter-clockwise for $h>0$. For any complex energy $E_B$ in the interior of the ellipse, the winding number $W(E_B)$ is 1 for $h>0$ and $-1$ for $h<0$, while when $E_B$ lies in the exterior of the ellipse one has $W(E_B)=0$. According to the bulk-edge correspondence, the interior of the ellipse (shaded area in the figure) corresponds to the energies of skin edge states under SIBC. Under OBC, the energy spectrum and corresponding skin modes collapse to the segment $(-\Delta,\Delta)$ on the real energy axis. (b) A topological interface obtained connecting two Hatano-Nelson chains with imaginary gauge fields $-h$ (for $n \leq 0$) and $h$ (for $n > 0$). The right panel shows the two overlapped ellipses, corresponding to the PBC energy spectra of the two bulk lattices, which are traveled in opposite directions. According to the bulk-edge correspondence, interface states do exist for any complex energy $E$ in the interior of the ellipse (shaded area in the figure), where $W_1(E) \neq W_2(E)$.}
\label{Fig1}
\end{figure}
is the complex energy of the skin mode. As the real parameters $k$ and $\mu$ are varied, the energies $E=E(k, \mu)$ of skin modes fill the interior of the ellipse. A main open question arises: is it possible to selectively prepare the system in one of the above topological skin modes? This question is not just merely concerning  the practical feasibility of exciting the system in the skin mode, rather it regards the true physical relevance of such skin edge states \cite{A1}. In fact, in a realistic one-dimensional system, such as in a photonic
lattice, the number of lattice sites $N$ is always finite and unavoidably open or periodic boundaries usually appear at the edges. Let us consider a finite lattice with sites from $n=-N_1$ to $n=N_2$, with integers $N_{1,2}>0$ possibly large and $N=N_1+N_2+1$. If we assume, for example, open boundary conditions (i.e. $\psi_n=0$ for $n>N_2$ and $n<-N_1$), in the thermodynamic limit $N_{1,2} \rightarrow \infty$ only a one-dimensional part of the interface skin states survive, which are picked out from the interface-state continuum, namely the edge states with real energies $E$ on the segment $(-\Delta, \Delta)$, corresponding to 
$\mu=0$ and $-\pi \leq k \leq \pi$. All other skin modes thus disappear in a finite-size system, although they can be observed for short times and therefore dubbed {\it quasi edge} skin modes in \cite{A1}. Examples of quasi edge skin states in a finite-size topological interface will be presented in Sec.4. Here we show that such quasi edge skin modes can survive, i.e. they can be stabilized, even in finite-sized systems provided that suitable on-site complex potentials, which supply or take energy from the system, are added at the edges of the chain. The main idea is that the skin edge state (\ref{palle}) is an {\it exact} eigenstate of the Hamiltonian in the finite-size system with open boundary conditions provided that the following on-site potentials are added 
\begin{equation}
V_{-N_1}= \Delta \exp(-ik - \mu) \; , \;\; V_{N_2}= \Delta \exp(ik + \mu). \label{poten}
\end{equation}
at the two edges $n=-N_1$ and $n=N_2$ of the lattice.
Such complex on-site potentials basically supply energy to the system on one edge (the one with positive imaginary part of the potential, corresponding to gain) and take energy from the system at the other edge (the one with negative imaginary part of the potential, corresponding to loss), effectively emulating an infinitely-extended topological interface. This simple observation suggests one that, after judicious tailoring of the complex on-site potentials at the edges of a finite-size topological interface, it might be possible to stabilize and somehow to selectively excite one among the various skin edge states of the infinitely-extended interface. The main result of this work, which is proven in the next section, is that the topological skin interface state (\ref{palle}) is obtained from the asymptotic dynamical evolution in the finite-size interface lattice with on-site potentials tailored according to Eq.(\ref{poten}), initially excited in the single edge site (either the left or right edge site), provided that ${\rm Im}(E)>0$. This means that, using a finite-size lattice with tailored complex potentials at the two edges, we can selectively excite any one of the topological skin interface state in the upper half area of the ellipse of Fig.1(b), where ${\rm Im}(E)>0$, but not the topological skin states with energies in the lower half area.

\section{Selective excitation of topological skin interface modes}
Let us consider a finite-size Hatano-Nelson topological interface with open boundary conditions, extended from $n=-N_1$ to $n=N_2$ , comprising $N=N_1+N_2+1$ sites with the topological interface located at $n=0$, i.e. with $h(n)=-h<0$ for $n \leq 0$ and $h(n)=h>0$ for $n>0$. We assume that complex on-site potentials $V_{-N_1}$ and $V_{N_2}$, defined by Eq.(\ref{poten}), are added at the two edges of the lattice at sites $n=-N_1$ and $n=N_2$, respectively. The evolution equations of the amplitudes $\psi_n(t)$ in the finite lattice then read
\begin{equation}
i \frac{d \psi_n}{dt}= \Delta \left\{ \exp [ h(n+1)] \psi_{n+1}+ \exp [-h(n)] \psi_{n-1} \right\} +(V_{-N_1} \delta_{n,-N_1}+V_{N_2}\delta_{n,N_2}  ) \psi_n(t)  \label{kepalle}
\end{equation}
($n=-N_1,-N_1+1,..., 0,1,..., N_2$) with the open boundary conditions $\psi_n(t)=0$ for $ n <-N_1$ and $n>N_2$. 
Without loss of generality, we can assume $\mu < 0$, i.e. $|V_{-N_1}|>\Delta$ and $|V_{N_2}|<\Delta$, as the other case $\mu \geq 0$ is simply obtained by reversing left and right edges of the lattice. The central result of this work is provided by the following:\\
{\it Theorem.} Let us indicate by $\psi_n(t)$ the evolution of the probability amplitudes in the finite-size lattice [Eq.(\ref{kepalle})] with the initial condition
\begin{equation}
\psi_n(0)=\delta_{n,-N_1},
\end{equation}
corresponding to the excitation of the left edge site of the lattice. Then $\psi_n(t)$ asymptotically evolves toward the topological skin mode (\ref{palle}), i.e. $\psi_n(t) \sim \exp(ikn+\mu n-h|n|-iEt)$ as $t \rightarrow \infty$,  provided that $-h<\mu <0$ and ${\rm Im}(E)>0$, where $E=2 \Delta \cosh(\mu+ik)$ is the complex energy of the topological skin mode.\\
In order to prove the above theorem, it is worth first introducing the non-unitary gauge transformation
\begin{equation}
\psi_n(t)= \left\{
\begin{array}{cc}
c_n(t) & n \leq 0 \\
c_n(t) \exp(-2hn) & n \geq 0
\end{array}
\right. \label{cazz}
\end{equation}
and to rescale the time variable $t$ so as $\Delta=1$.
 The non-unitary gauge transformation (\ref{cazz}) basically eliminates the topological interface, making the imaginary gauge field uniform all along the finite-size  lattice. In fact, under the transformation (\ref{cazz}) and with $\Delta=1$ the wave amplitudes $c_n(t)$ satisfy the following coupled equations
 \begin{equation}
 i \frac{dc_n}{dt}=\exp(-h) c_{n+1}+\exp(h) c_{n-1}+(V_{-N_1} \delta_{n,-N_1}+V_{N_2} \delta_{n,N_2}) c_n \label{figa}
 \end{equation}
 ($n=-N_1, -N_1+1,...,0,1,...N_2$) with the open boundary conditions $c_n(t)=0$ for $n<-N_1$, $n>N_2$ and with the initial condition 
 \begin{equation}
 c_n(0)=\delta_{n,-N_1}. \label{basta}
 \end{equation}
  Note that, after the gauge transformation (\ref{cazz}), the skin interface state (\ref{palle}) takes the form
  \begin{equation}
 c_n^{(skin)}(t)= A \exp[ (ik+\mu +h) (n+N_1)-iEt] \label{skino}
 \end{equation}
 so that to prove the theorem we need to show that
 \begin{equation}
 \lim_{t \rightarrow \infty} \epsilon(t)=0
 \end{equation}
  for a suitable amplitude $A$, where $\epsilon(t)$ is the deviation function defined by
  \begin{equation}
  \epsilon(t) \equiv \frac{\| c_n(t)-c_n^{(skin)}(t) \|^2}{\| c_n^{(skin)}(t) \|^2}= \frac{\sum_{n=-N_1}^{N_2} |c_n(t)-c_n^{(skin)}(t)|^2}{ \sum_{n=-N_1}^{N_2} |c_n^{(skin)}(t)|^2}. \label{error}
  \end{equation}
 The analytical solution $c_n(t)$ to Eqs.(\ref{figa}) with the initial condition (\ref{basta}) can be obtained using the Laplace transform method (see, for instance, \cite{Laplace}). After introduction of the Laplace transform $\hat{c}_n(s)$ of the wave amplitude $c_n(t)$
 \begin{equation}
 \hat{c}_n(s)=\int_0^{\infty}dt c_n(t) \exp(-st)
 \end{equation}
 with ${\rm Re}(s)> \eta>0$, from Eqs.(\ref{figa}) and (\ref{basta}) one obtains
 \begin{equation}
 (is-V_{-N_1} \delta_{n,-N_1}-V_{N_2} \delta_{n,N_2}) \hat{c}_n(s)-t_1 \hat{c}_{n+1}(s)-t_2 \hat{c}_{n-1(s)}=i \delta_{n,-N_1} \label{sistema}
 \end{equation}
 where we have set 
\begin{equation}
 t_1 \equiv \exp(-h) \; ,\;\;\; t_2 \equiv \exp(h).
 \end{equation}
Equation (\ref{sistema}) is a linear algebraic system, that can be solved for $\hat{c}_{n}(s)$ ($-N_1 \leq n \leq N_2$) by the Cramer$^{\prime}$s rule. As shown below, for our purposes it is enough to calculate $\hat{c}_{-N_1}(s)$, which takes the form
\begin{equation}
\hat{c}_{-N_1}(s)=\frac{i}{is-V_{-N_1}-\Sigma_N(s)}
\end{equation}
where the self-energy $\Sigma_N(s)$ is derived in Appendix A and reads
\begin{equation}
\Sigma_N(s)=\frac{(is-V_{N_2}) \sin [(N-1) \theta ]- \sin [(N-2) \theta] }{(is-V_{N_2}) \sin (N \theta) -\sin [(N-1) \theta]}. \label{self}
\end{equation}
In the above equation, we introduced the complex angle $\theta$ via the relation
\begin{equation}
2 \cos \theta = i s \label{barba}.
\end{equation}
 Once $\hat{c}_{-N_1}(s)$ has been found, the corresponding wave amplitude $c_{-N_1}(t)$ is obtained after inversion as a contour integral in complex $s$ plane (Bromwich integral) 
\begin{equation}
c_{-N_1}(t)=\frac{1}{2 \pi i} \int_{\rm B} ds \; \exp(st) \hat{c}_{-N_1}(s) \label{inversion}
\end{equation}
where the Bromwich path B is the horizontal contour ${\rm Re}(s)=\eta$ in the complex plane and $\eta>0$ is chosen so that all singularities of $\hat{c}_n(s)$ are below B (Fig.2).\\
 To calculate the 
Bromwich integral, let us first consider the $N \rightarrow \infty$ limit. Assuming the additional constraint ${\rm Im} (\theta)>0$, the self-energy $\Sigma_N(s)$ converges to the simple function 
\begin{equation}
\Sigma_{\infty}(s)=\exp(i \theta) \label{self1}
\end{equation}
and thus
\begin{equation}
\hat{c}_{-N_1}(s)=\frac{i}{is-V_{-N_1}-\Sigma_{\infty}(s)}=\frac{i}{\exp(-i \theta)-V_{-N_1}}. \label{noia}
\end{equation}
Note that, owing to the constraint ${\rm Im} (\theta)>0$, the self-energy $\Sigma_{\infty}(s)$ shows a branch cut on the segment I of the imaginary axis defined by $s=i \omega$, with $-2 \leq \omega \leq 2$ [Fig.2(a)]. 
From Eqs.(\ref{barba}) and (\ref{noia}), taking into account that $V_{-N_1}=\exp(-ik-\mu)$ and that $ \mu <0$, it readily follows that $c_{-N_1}(s)$ has a single pole at 
\begin{equation}
s=s_p=-i E, \label{pol}
\end{equation}   
where $E=2 \cosh(\mu+ik)$. Provided that ${\rm Im}(E)>0$, the pole lies above the branch cut. In this case we can compute the Bromwich integral by deforming the contour B as shown in Fig.2(a), so that one has
\begin{equation}
c_{-N_1}(t)=\frac{1}{2 \pi i} \int_{\rm H} ds \; \exp(st) \hat{c}_{-N_1}(s) + \frac{1}{2 \pi i} \int_{\sigma} ds \; \exp(st) \hat{c}_{-N_1}(s) \label{inversion2}
\end{equation}
where the Hankel path H encircles the branch cut I while $\sigma$ encircles the pole at $s=s_p$. Since ${\rm Re}(s_p)>0$, the latter contribution to the integral dominates over the first contribution for long times, and can be readily computed from the residue of $\hat{c}_{-N_1}(s)$ at $s=s_p$. This yields
\begin{equation}
c_{-N_1}(t) \sim A \exp(-i Et) \label{appro}
\end{equation}
where 
\begin{equation}
%A=\frac{4}{2-\frac{s_p}{\sqrt{1+(s_p/2)^2}}}. \label{ampli}
A=1-\frac{1}{V_{-N_1}^2} \label{ampli}
\end{equation}
The Hankel path contribution provides a correction to Eq.(\ref{appro}), however it remains bounded as $ t \rightarrow \infty$ and so it can be neglected after an initial transient. Once we have computed the asymptotic behavior of $c_{-N_1}(t)$ at long times, we can directly calculate the asymptotic form of all other amplitudes $c_n(t)$ from Eqs.(\ref{figa}) and (\ref{appro}) by an iterative procedure using the relations
\begin{eqnarray}
c_{n+1}(t)  & = & i \exp(h) \frac{dc_n}{dt}- V_{-N_1}\exp(h)  c_n(t) \;\;\;\;\;\;\; (n=-N_1) \\
c_{n+1}(t)  & = & - \exp(2h) c_{n-1}(t)+i \exp(h) \frac{dc_n}{dt} \;\;\;\;\;\; (n>-N_1)
\end{eqnarray}
This yields
\begin{equation}
c_n(t) \sim A \exp [(ik+\mu+h)(n+N_1)-iEt] \label{ohh}
\end{equation}
which is valid in the long time limit. From a comparison of Eqs.(\ref{skino}) and (\ref{ohh}) it then readily follows that $\lim_{t \rightarrow \infty} \epsilon(t)=0$, which proves the theorem in the $N \rightarrow \infty$ limit.\\
\begin{figure}[!h]
\centering\includegraphics[width=5in]{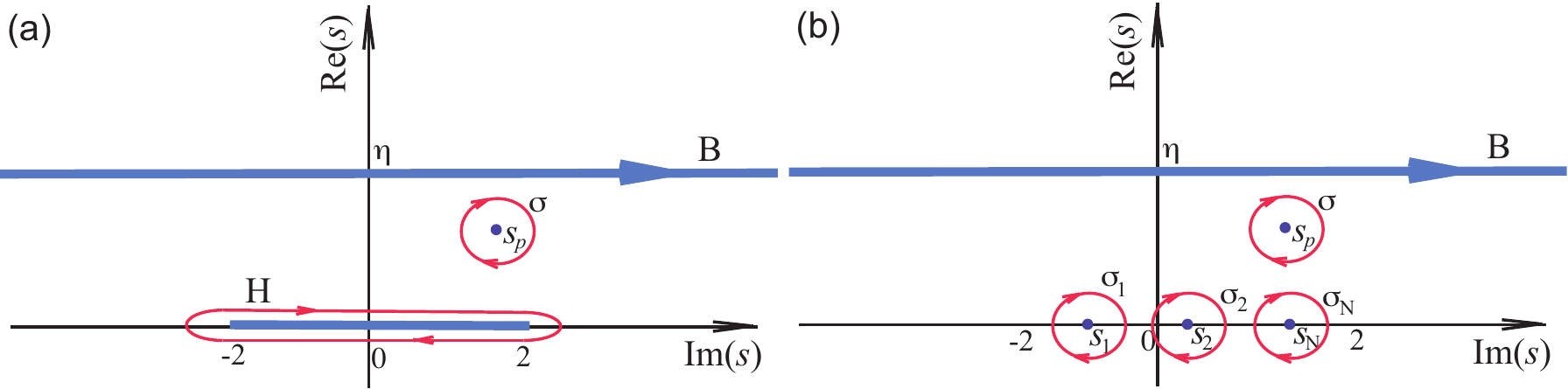}
%%% where xxxxxx name represents "figurename.eps"
\caption{Integration paths in complex $s$ plane used for the calculation of the temporal behavior of the amplitude $c_{-N_1}(t)$ at the left edge site of the topological interface. The horizontal bold line is the Bromwich path B with $ {\rm Re}(s)= \eta$ and $\eta>0$ large enough such that all the singularities of $\hat{c}_{-N_1}(s)$ lie below the line B. (a) The case of a topological interface in the $N \rightarrow \infty$ limit.  In this case $\hat{c}_{-N_1}(s)$ shows a single pole at $s=s_p=-iE$, which lies above the imaginary axis provided that ${\rm Im}(E)>0$, and a branch cut along the segment I on the imaginary axis, from $s=-2i$ to $s=2i$ (solid bold segment in the figure). The Bromwich path can be deformed to $\sigma \bigcup {\rm H}$, where $\sigma$ encircles the pole $s=s_p$ whereas the Hankel path H encircles the branch cut. (b) The case of a topological interface with a finite number $N$ of lattice sites. In this case $\hat{c}_{-N_1}(s)$ shows $(N+1)$ poles and no branch cuts. The dominant pole is at $s=s_p=-iE$, whereas the other $N$ poles $s_l=-2 i \cos [\pi l /(N+1)]$ ($l=1,2,3,...,N$) lie on the imaginary axis, between $s=-2i$ and $s=2i$. The Bromwich path can be deformed to the set of closed loops that encircle the various poles.}
\label{Fig2}
\end{figure}
 In a topological interface with finite $N$, a similar analysis can be performed. In this case the self-energy $\Sigma_N(s)$ is given by Eq.(\ref{self}) and after some straightforward calculations $\hat{c}_{-N_1}(s)$ takes the form
 \begin{equation}
 \hat{c}_{-N_1}(s)=\frac{ \sin [(N+1) \theta]-V_{N_2} \sin (N \theta)}{(s-s_p) \sin [(N+1) \theta]}
 \end{equation}
 where $s_p$ is given by Eq.(\ref{pol}). Clearly, $\hat{c}_{-N_1}(s)$ shows $(N+1)$ poles: one pole is located exactly at $s=s_p=-iE$, like in the $N \rightarrow \infty$ limit, whereas the other $N$ poles 
 are obtained at the angles $\theta_l= l \pi/(N+1)$, i.e. $s_l=-2 i \cos \theta_l= -2 i \cos [ l \pi /(N+1)]$ ($l=1,2,3,..,N$). Note that such additional poles
  lie on the segment I of the imaginary axis, and become dense on I as $N \rightarrow \infty$. In other words, the branch cut I of $\hat{c}_{-N_1}(s)$ in the $N \rightarrow \infty$ limit is replaced for finite $N$ by a dense set of poles on I, as shown in Fig.2(b). The Bromwich integral is then given by the sum of the residues arising from the $(N+1)$ poles. Assuming ${\rm Im}(E)>0$, the dominant pole is the one at $s=s_p=-iE$, and thus after an initial transient the asymptotic form of $c_{-N_1}(t)$ is again given by Eqs.(\ref{appro}). Once the asymptotic form of $c_{-N_1}(t)$ has been determined, the asymptotic form of the other amplitudes $c_n(t)$ is obtained from the recursive relations
\begin{eqnarray}
c_{n+1}(t)  & = & i \exp(h) \frac{dc_n}{dt}- V_{-N_1}\exp(h)  c_n(t) \;\;\;\;\;\;\; (n=-N_1) \\
c_{n+1}(t)  & = & - \exp(2h) c_{n-1}(t)+i \exp(h) \frac{dc_n}{dt} \;\;\;\;\;\; (-N_1<n \leq N_2-1) \\
%c_{n+1}(t)  & = & - \exp(2h) c_{n-1}(t)+i \exp(h) \frac{dc_n}{dt} -V_{N_2} \exp(h)c_n(t) \;\;\; (n=N_2-1).
\end{eqnarray}
This yields again the asymptotic form Eq.(\ref{ohh}) for $c_n(t)$, thus proving the theorem for finite $N$ as well. 
%%% Numbered equation
%\begin{align}\label{1.1}
%\begin{split}
%\frac{\partial u(t,x)}{\partial t} &= Au(t,x) \left(1-\frac{u(t,x)}{K}\right)-B\frac{u(t-\tau,x) w(t,x)}{1+Eu(t-\tau,x)},\\
%\frac{\partial w(t,x)}{\partial t} &=\delta \frac{\partial^2w(t,x)}{\partial x^2}-Cw(t,x)+D\frac{u(t-\tau,x)w(t,x)}{1+Eu(t-\tau,x)},
%\end{split}
%\end{align}

%\begin{align}\label{1.2}
%\begin{split}
%\frac{dU}{dt} &=\alpha U(t)(\gamma -U(t))-\frac{U(t-\tau)W(t)}{1+U(t-\tau)},\\
%\frac{dW}{dt} &=-W(t)+\beta\frac{U(t-\tau)W(t)}{1+U(t-\tau)}.
%\end{split}
%\end{align}

%%%% Unnumbered equation
%\begin{eqnarray}
%\frac{\partial(F_1,F_2)}{\partial(c,\omega)}_{(c_0,\omega_0)} = \left|
%\begin{array}{ll}
%\frac{\partial F_1}{\partial c} &\frac{\partial F_1}{\partial \omega} \\\noalign{\vskip3pt}
%\frac{\partial F_2}{\partial c}&\frac{\partial F_2}{\partial \omega}
%\end{array}\right|_{(c_0,\omega_0)}\notag\\
%=-4c_0q\omega_0 -4c_0\omega_0p^2 =-4c_0\omega_0(q+p^2)>0.
%\end{eqnarray}

\begin{figure}[!h]
\centering\includegraphics[width=5in]{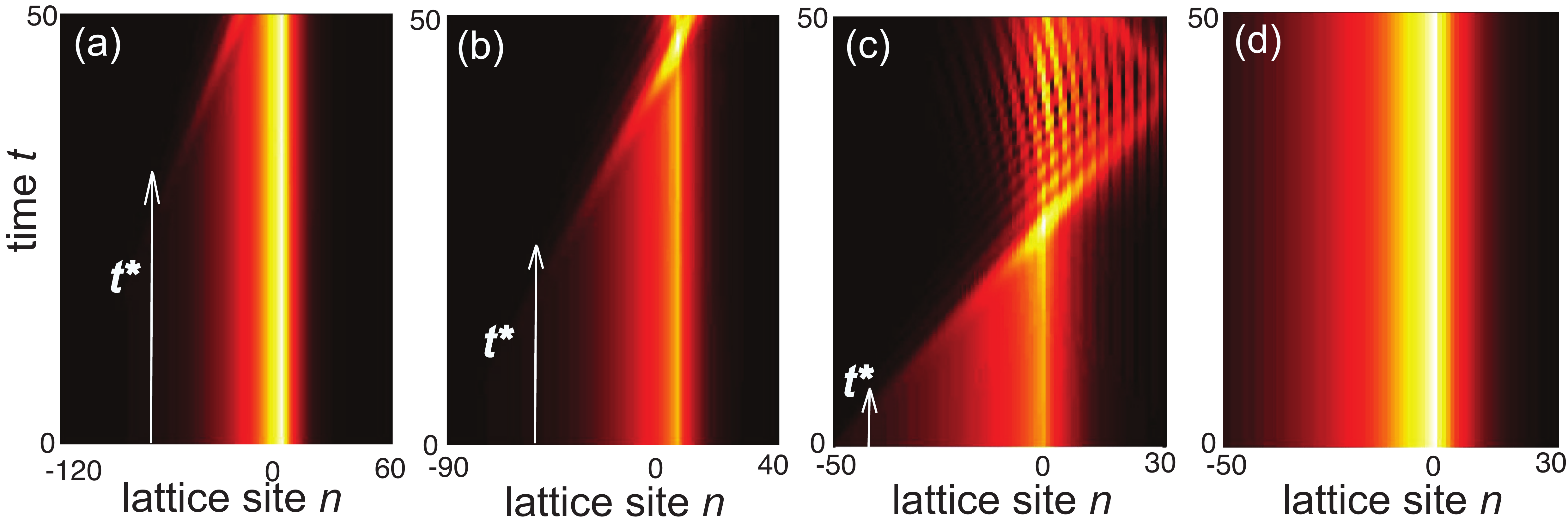}
%%% where xxxxxx name represents "figurename.eps"
\caption{Quasi edge (metastable) topological interface states in a finite-size Hatano-Nelson interface with open boundary conditions for parameter values $\Delta=1$ and $h=0.1$. The system is initially excited in the skin interface state $\psi_n(0)=\exp(ikn+\mu n -h |n|)$ with $\mu=-0.05$ and $k=-\pi/3$. The panels show on a pseudo color map the temporal evolution of the normalized amplitudes $|\psi_n(t)| / \sqrt{\sum_n | \psi_n(t)|^2}$.
In (a-c) the edge on-site potentials $V_{-N_1}$ and $V_{N_2}$ vanish, so as the topological skin state survives only for a time $t^*$, after which it is fully destroyed owing to finite lattice size effects. The time scale $t^*$ increases as the system size $N_{1,2}$ increases [$N_1=120, N_2=60$ in (a); $N_1=90, N_2=40$ in (b); $N_1=50, N_2=30$ in (c)]. As soon as the appropriate potentials $V_{-N_1}= \exp(-ik - \mu)$ and $V_{N_2}=1/ V_{-N_1}$ are added at the left and right edge sites of the lattice, the topological skin state is stabilized, as shown in panel (d).}
\label{Fig3}
\end{figure}

\begin{figure}[!h]
\centering\includegraphics[width=5in]{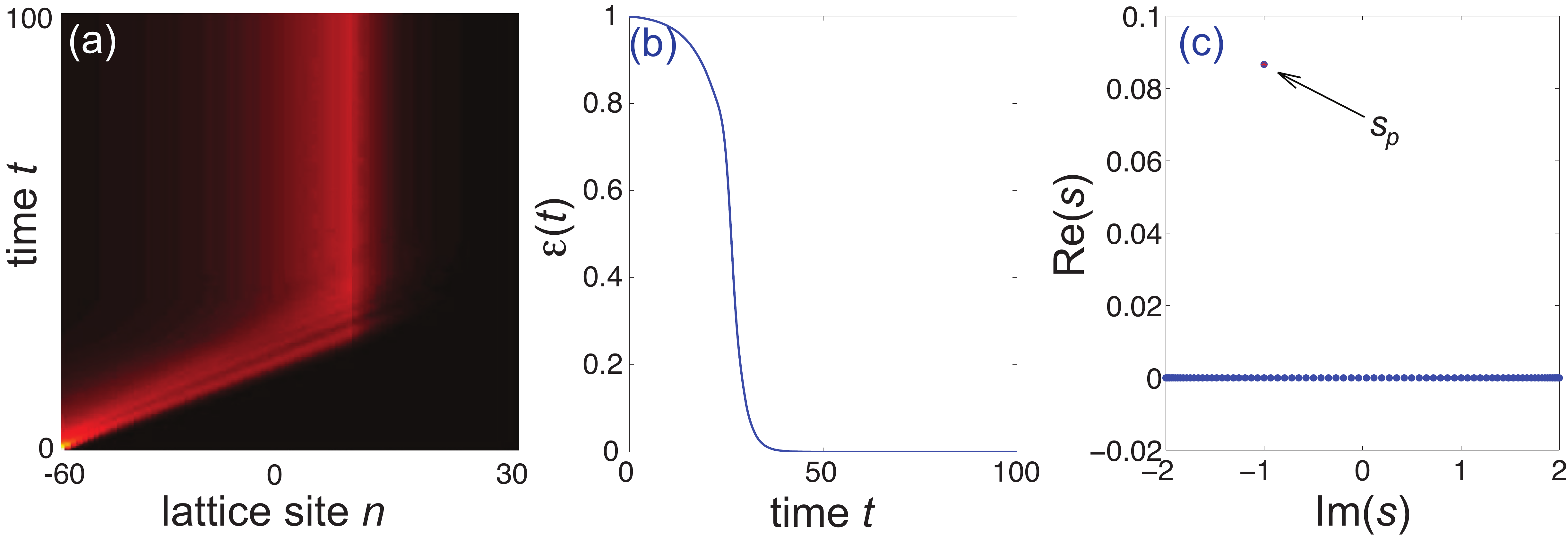}
%%% where xxxxxx name represents "figurename.eps"
\caption{Dynamical formation of the topological interface skin state in a finite Hatano-Nelson lattice with open boundary conditions for parameter values $\Delta=1$ and $h=0.1$. The system is initially excited at the left edge site, and on-site potentials $V_{-N_1}= \exp(-ik - \mu)$ and $V_{N_2}=1/ V_{-N_1}$ are added at the left and right edge sites of the lattice ($\mu=-0.05$, $k= - \pi/3$). (a) Temporal evolution on a pseudo color map  of the normalized amplitudes $|\psi_n(t)| / \sqrt{\sum_n | \psi_n(t)|^2}$. (b) Behavior of the deviation function $\epsilon(t)$. (c) Poles of $\hat{c}_{-N_1}(s)$ (circles) in the complex $s$ plane. The topological skin mode arises from the dominant pole $s_p=-iE= 0.0866 - 1.0013i$ of $\hat{c}_{-N_1}(s)$.}
\label{Fig4}
\end{figure}

\begin{figure}[!h]
\centering\includegraphics[width=5in]{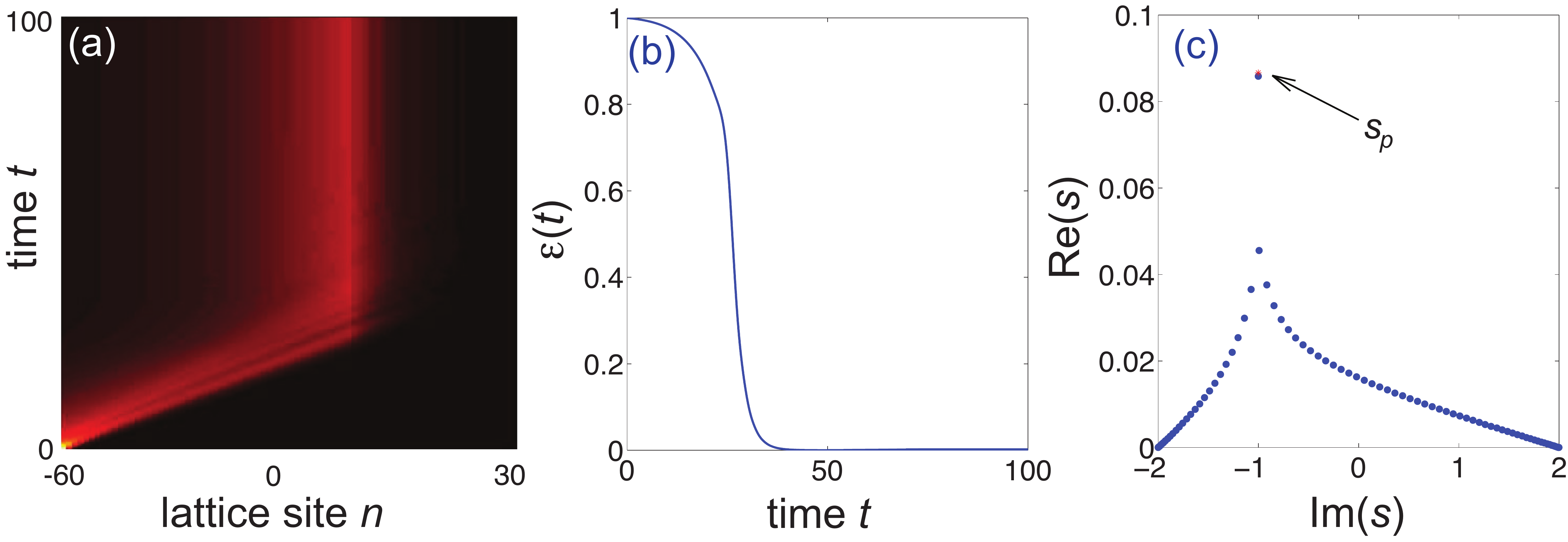}
%%% where xxxxxx name represents "figurename.eps"
\caption{Same as Fig.4, but for $V_{-N_2}=0$.}
\label{Fig5}
\end{figure}

\section{Numerical simulations}
We checked the predictions of the theoretical analysis by extended numerical simulations of the coupled equations (\ref{kepalle}) in time domain using an accurate fourth-order variable-step Runge-Kutta method. Some illustrative examples are shown in Figs.3,4 and 5 and 6. In the simulations of Figs.3,4, and 5 we assumed the parameter values $\Delta=1$, $h=0.1$, $\mu=-0.05$ and $k=-\pi/3$, corresponding to the eigenenergy $E=2 \Delta \cosh (\mu+ik)=1.0013 + 0.0866i$ of the skin interface state and to the on-site potentials $V_{-N_1}= 0.5256 + 0.9104i$, $V_{N_2}=1/V_{N_1}=0.4756 - 0.8238i$ at the left and right lattice edges, respectively. Note that, since ${\rm Im}(V_{-N_1})>0$ (gain) and ${\rm Im}(V_{N_2})<0$ (loss), energy is basically supplied to the lattice at the left edge site $n=-N_1$, while it is taken from the lattice at the right edge site $n=N_2$. 
Figures 3(a-c) illustrate the concept of quasi edge skin modes \cite{A1}, i.e. the metastability of the topological interface skin state in a finite lattice for a few increasing values of $N_{1,2}$ when we set $V_{-N_1}=V_{N_2}=0$. The results are obtained integrating Eq.(\ref{kepalle}) with $V_{-N_1}=V_{N_2}=0$ and assuming as an initial condition $\psi_n(t=0)$ the topological skin state, defined by Eq.(\ref{palle}), cutted at the edges. The figures clearly show that, for $V_{-N_1}=V_{N_2}=0$ the dynamics  
looks just like the one of an eigenstate up to some time $t^*$, above which the state is completely destroyed owing to edge effects. The survival time $t^*$ of the quasi eigenstate increases
with the system size $N_{1,2}$, according to the analysis of Ref.\cite{A1}. Conversely, when we add the appropriate on-site potentials $V_{-N_1}$ and $V_{N_2}$ at the lattice edges the state survives for ever, as shown in Fig.3(d), since the additional on-site potentials basically emulate the infinite lattice limit. The most important result provided by the theorem stated in Sec.3 is that the skin interface state can be dynamically generated by initial single-site excitation of the lattice at the left edge site. This is shown in Fig.4, where the formation and stabilization  of the topological interface skin mode is clearly demonstrated, after an initial time transient, when the on-site potentials $V_{-N_1}$ and $V_{N_2}$ are added at the lattice edges. In particular, according to the theoretical predictions the deviation function $\epsilon(t)$, defined by Eq.(\ref{error}), decays toward zero, indicating the convergence of $\psi_n(t)$ to the topological skin eigenstate $\psi_n^{(skin)}(t)$. As we change the values $V_1= \exp(-ik - \mu)$ and $V_2=1/V_1$ so as to meet the conditions of the menn theorem stated in previous section, i.e. $-h<\mu<0$ and ${\rm Im}(E)>0$, we can selectively generate and tune the dynamically-generated topological skin mode. 
 Interestingly, in some cases the skin edge state can be generated even when the on-site potential is added to the left edge site, supplying energy (gain) to the lattice, but not at the right edge, i.e. by assuming $V_{-N_1}=\exp(-ik-\mu)$ and $V_{N_2}=0$. This is shown in Fig.5. In this case, the location of the $(N+1)$ poles of the Laplace transform $\hat{c}_{-N_1}(s)$, shown in Fig.5(c), clearly deviates from the distribution predicted by the theoretical analysis [Fig.2(b) and Fig.4(c)]; in particular the dense poles on the imaginary $s$ axis shift toward the ${\rm Re}(s)>0$ half plane, however a dominant pole very close to the theoretical value $s_p=-iE$ survives. This explains why, even though $V_{-N_2}=0$, the topological interface skin mode can be dynamically generated after an initial transient, as a result of the dominant pole contribution to $\hat{c}_{-N_1}$. It should be noted that such a result, i.e. excitation of the topological skin interface mode with only the left on-site complex potential, is very appealing from a practical viewpoint, since it requires just to tune the potential (gain and on-site energy offset) of one site of the lattice. However, this is not possible in all cases, since it requires that the perturbation $V_{N_2}=0$ does not  substantially change the position of dominant poles of $\hat{c}_{-N_1}(s)$. An example is shown in Figs.6 and 7, where in this case stabilization of the topological skin mode strictly requires both on-site potentials at the two edges.
\begin{figure}[!h]
\centering\includegraphics[width=5in]{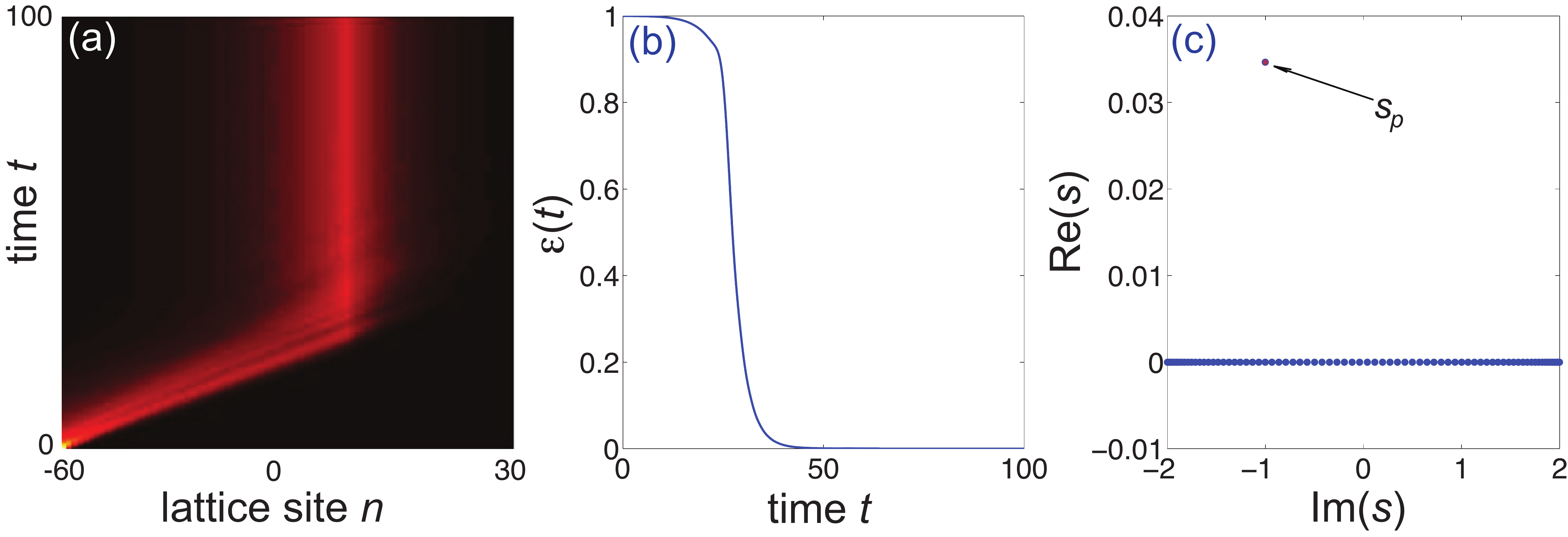}
%%% where xxxxxx name represents "figurename.eps"
\caption{Same as Fig.4, but for $\mu=-0.02$ and $k= - \pi/3$.}
\label{Fig6}
\end{figure}
\begin{figure}[!h]
\centering\includegraphics[width=5in]{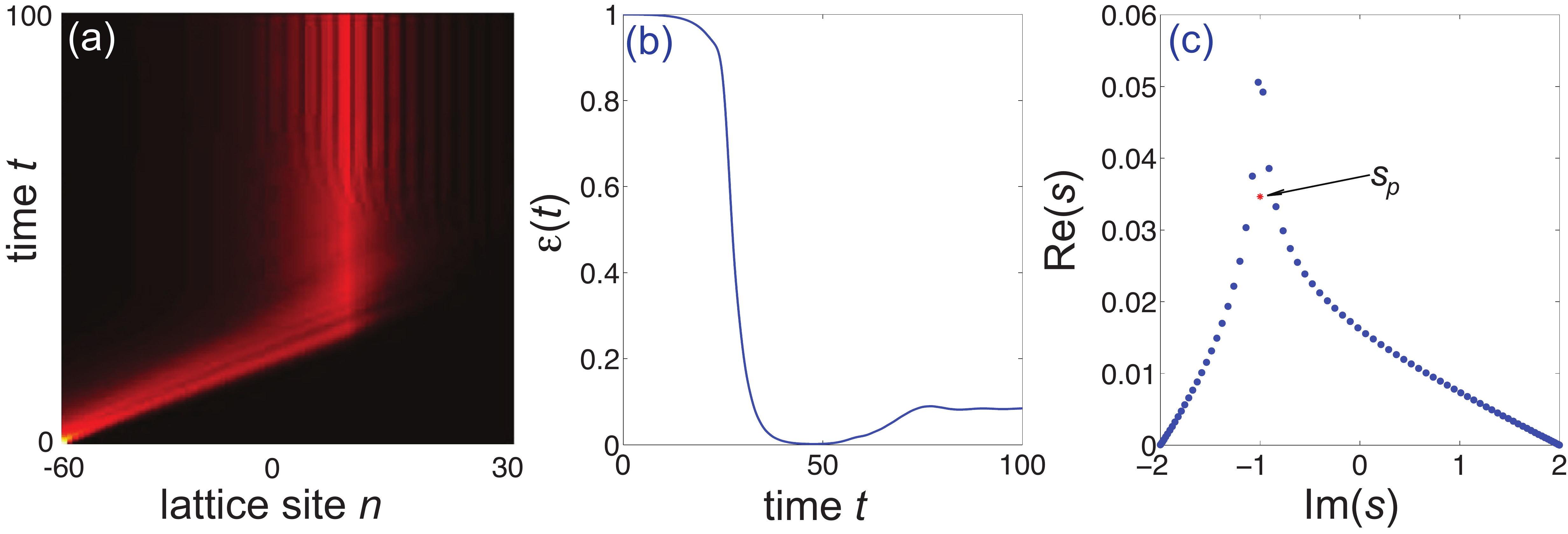}
%%% where xxxxxx name represents "figurename.eps"
\caption{Same as Fig.5, but for $\mu=-0.02$ and $k= - \pi/3$. Note that in this case the vanishing of $V_{N_2}$ largely perturbs the position of poles of $c_{-N_1}(s)$ in the complex $s$ plane, and the topological skin interface state cannot be dynamically generated. In panel (c) $s_p=-iE$ is the position of the dominant pole predicted by the theoretical analysis in the ideal case $V_{N_2}=1/V_{-N_1}$.}
\label{Fig7}
\end{figure}

%\vspace*{-5pt}

%\noindent The output for table is:\vspace*{-7pt}

%\begin{table}[!h]
%\caption{An Example of a Table}%%%Table caption goes here
%\label{table_example}
%\begin{tabular}{llll}%%%The number of columns has to be defined here
%\hline
%date &Dutch policy &date &European policy \\
%\hline
%1988 &Memorandum Prevention &1985 &European Directive (85/339) \\
%1991--1997 &{\bf Packaging Covenant I} & & \\
%1994 &Law Environmental Management &1994 &European Directive (94/62) \\
%1997 &Agreement Packaging and Packaging Waste & & \\\hline
%\end{tabular}
%\vspace*{-4pt}
%\end{table}%%%End of the table

\section{Conclusion}
To conclude, in this work we suggested a route toward the selective excitation of topological skin edge states in a finite-size lattice, based on a judicious tailoring of the complex on-site potentials at the lattice edges. Remarkably, the skin edge state is dynamically generated by single-site edge excitation of the lattice after an initial transient. Our results indicate that a double-continuum set of skin edge states can be stabilized by the edge on-site potentials, preventing their disruption that one would observe in a finite lattice geometry \cite{A1}. The method has been illustrated by considering skin modes at a topological Hatano-Nelson interface, which enables for a full analytical treatment, however the stabilization strategy should be feasible for an extension to other non-Hermitian lattice models: the basic idea is to emulate a semi-infinite geometry in a finite-size lattice by adding suitable complex on-site potentials at the truncated edge. 
The ability of generating and stabilizing skin edge states sheds new light onto the physical relevance and robustness of such states, and could be of potential relevance in future sophisticated applications of non-Hermitian skin modes. For example, skin modes enjoy the fantastic property of being self-healing waves, i.e. the can self-reconstruct their shape after being scattered off by an obstacle \cite{unpub}. Clearly, to demonstrate and harness such a remarkable property in any application one should be able to stably generate such skin states.

\vskip6pt

\enlargethispage{20pt}

\ethics{There are not ethical issues for this paper.}

\dataccess{This article has no additional data}
%Insert data accessibility text here. '(If no information, then please include the text ``This article has no additional data'').

%\aucontribute{Insert author contributions text here (to be included if more than one author).}

\competing{There are not competing interests.}

\funding{No fundings for this paper.}

\ack{ The author acknowledges the Spanish
State Research Agency, through the Severo-Ochoa and Maria de
Maeztu Program for Centers and Units of Excellence in R\&D
(Grant No. MDM-2017-0711).}

\disclaimer{No disclaimer for this paper.}

%%%%%%%%%% Insert bibliography here %%%%%%%%%%%%%%

\vskip2pc

%\noindent {\bf Please follow the coding for references as shown below.}

\appendix
\section{Derivation of the self-energy}
The Laplace transform $\hat{c}_{-N_1}(s)$ of the wave amplitude $c_{-N_1}(t)$ at the edge site $n=-N_1$of the lattice is obtained from the solution of the linear system Eq.(\ref{sistema}). Using the Cramer$^{\prime}$s rule, 
one can write
\begin{equation}
\hat{c}_{-N_1}(s)= \frac{iP_N(s)}{(is-V_{-N_1})P_N(s)-t_1 t_2 P_{N-1}(s)}=\frac{i}{(is-V_{-N_1})-t_1 t_2 P_{N-1}(s)/P_N(s)}
\end{equation}
where $P_{N}(s)$ is $N \times N$ determinant
\begin{equation}
P_N(s)= 
\left|  
\begin{array}{ccccccccccc}
 is & -t_1 & 0 & 0 & 0 &... & 0 & 0 & 0 & 0 & 0 \\
 -t_2 & is & -t_1 & 0 & 0 & ... & 0 & 0 & 0 & 0 & 0 \\
0 & -t_2 &   is & -t_1 & 0 & ... & 0 & 0 & 0 & 0 & 0 \\
 ... & ... & ... & ... & ... &... & ... & ... & ... & ... & ... \\
  0 & 0 & 0 & 0 & 0 & 0 & ... & -t_2 & is & -t_1 & 0 \\
  0 & 0 & 0 & 0 & 0 & 0 & ... & 0 & -t_2 & is & -t_1 \\
  0 & 0 & 0 & 0 & 0 & 0 & ... & 0 & 0 & -t_2 & is-V_{N_2} \\
\end{array}
\right|.
\end{equation}
Taking into account that $t_1 t_2=1$, one obtains
\begin{equation}
\hat{c}_{-N_1}(s)=\frac{i}{is-V_{-N_1}-\Sigma_N(s)}
\end{equation}
where the self-energy is defined by
\begin{equation}
\Sigma_N(s) = \frac{P_{N-1}(s)}{P_N(s)}. \label{raffa0}
\end{equation}
To calculate the self-energy, we need to compute the determinant $P_N(s)$.
Expanding the determinant from the last row, it can be readily shown that
\begin{equation}
P_N(s)=(is-V_{N_2})Q_{N-1}-t_1t_2 Q_{N-2}(s) \label{raffa1}
\end{equation}
where $Q_N(s)$ is the determinant of a $N \times N$ tridiagonal Toeplitz matrix, namely
\begin{equation}
Q_N(s)= 
\left|  
\begin{array}{ccccccccccc}
 is & -t_1 & 0 & 0 & 0 &... & 0 & 0 & 0 & 0 & 0 \\
 -t_2 & is & -t_1 & 0 & 0 & ... & 0 & 0 & 0 & 0 & 0 \\
0 & -t_2 &   is & -t_1 & 0 & ... & 0 & 0 & 0 & 0 & 0 \\
 ... & ... & ... & ... & ... &... & ... & ... & ... & ... & ... \\
  0 & 0 & 0 & 0 & 0 & 0 & ... & -t_2 & is & -t_1 & 0 \\
  0 & 0 & 0 & 0 & 0 & 0 & ... & 0 & -t_2 & is & -t_1 \\
    0 & 0 & 0 & 0 & 0 & 0 & ... & 0 & 0 & -t_2 & is \\
\end{array}
\right|.
\end{equation}
The determinant $Q_N(s)$ can be readily calculated from the recursive relation  $Q_N(s)=is Q_{N-1}(s)-t_1t_2 Q_{N-2}(s)$ with $Q_0(s)=1$, $Q_1(s)=is$. Taking into account that $t_1t_2=1$ one obtains
\begin{equation}
Q_N(s)= \frac{\sin (N+1) \theta}{\sin \theta} \label{raffa2}
\end{equation}
where we introduced the complex angle $\theta$ via the relation
\begin{equation}
2 \cos \theta= is.
\end{equation}
Substitution of Eq.(\ref{raffa2}) into Eq.(\ref{raffa1}), one finally obtains the following form of the self-energy $\Sigma_N(s)$ [Eq.(\ref{raffa0})]
\begin{equation}
\Sigma_N(s)=\frac{(is-V_{N_2}) \sin [(N-1) \theta ]- \sin [(N-2) \theta] }{(is-V_{N_2}) \sin (N \theta) -\sin [(N-1) \theta]}. 
\end{equation}

%\noindent If maintaining .bib file for references, then please use "RS.bst" to generate the references.

%\noindent Example:

%\verb+\bibliographystyle{RS}+ %%%% .BST file

%\verb+\bibliography{sample}+ %%%%% .Bib file

\end{document}